\documentclass[largeformat]{interact}

\usepackage{graphicx}
\usepackage{dcolumn}
\usepackage{bm}
\usepackage{color}

\begin{document}

\newcommand{\Orf}[0]{\Omega_\textrm{rf}} 
\newcommand{\kB}[0]{k_\textrm{B}} 


\title{Experimental apparatus for overlapping a ground-state cooled ion with ultracold atoms}

\author{
\name{Ziv Meir\textsuperscript{a}, Tomas Sikorsky\textsuperscript{a}, Ruti Ben-shlomi\textsuperscript{a}, Nitzan Akerman\textsuperscript{a}, Meirav Pinkas\textsuperscript{a}, Yehonatan Dallal\textsuperscript{a} and Roee Ozeri\textsuperscript{a}}
\affil{\textsuperscript{a}Department of Physics of Complex Systems, Weizmann Institute of Science, Rehovot 7610001, Israel}
}

\maketitle


\begin{abstract}
Experimental realizations of charged ions and neutral atoms in overlapping traps are gaining increasing interest due to their wide research application ranging from chemistry at the quantum level to quantum simulations of solid-state systems. 
Here, we describe a system in which we overlap a single ground-state cooled ion trapped in a linear Paul trap with a cloud of ultracold atoms such that both constituents are in the $\mu$K regime. 
Excess micromotion (EMM) currently limits atom-ion interaction energy to the mK energy scale and above. We demonstrate spectroscopy methods and compensation techniques which characterize and reduce the ion's parasitic EMM energy to the $\mu$K regime even for ion crystals of several ions.
We give a substantial review on the non-equilibrium dynamics which governs atom-ion systems. The non-equilibrium dynamics is manifested by a power-law distribution of the ion's energy. 
We overview the coherent and non-coherent thermometry tools which we used to characterize the ion's energy distribution after single to many atom-ion collisions.
\end{abstract}

\tableofcontents

\section{\label{sec:Introduction}Introduction}
The increased appeal of hybrid system stems from the potential benefits of combining the strengths of its composing systems. Trapped ions are leading candidates for the realization of quantum computers \cite{Kielpinski2002,Linke2017} and they perform outstandingly as local probes \cite{Kotler2013,Baumgart2016,Shaniv2017,Ruster2017} and for quantum meteorology applications \cite{Rosenband2008,Kotler2011,Kotler2014,Ludlow2015}. Ultracold clouds of neutral atoms are routinely used for quantum simulations of tailored Hamiltonians and the study of quantum many-body physics \cite{cornell2002nobel,bloch2012quantum}. In the last two decades, hybrid atom-ion systems \cite{Harter2014,Sias2014,willitsch2014ion} were realized in several laboratories around the world \cite{Makarov2003,JianLin2005,Grier2008,Zipkes2010,Schmid2010,Rellergert2011,Hall2011,Sivarajah2012,Ravi2012,Haze2013,Meir2016} with exciting prospects such as emulating solid-state systems \cite{Bissbort2013}, implementing double-well bosonic Josephson junctions \cite{Gerritsma2012} and performing atom-ion quantum gates \cite{Doerk2010}.

The interaction between atoms and ions is governed by the polarization potential which scales asymptotically as $-1/r^4$ and extends to 100's of nm. At the atomic scale the molecular energy surfaces exhibit vast complexity giving control of chemical reactions and spin dynamics at the single particle level \cite{Ratschbacher2012,Hall2013,Harter2013,Ratschbacher2013}. Purely quantum collisional dynamics is expected when the de Broglie wavelength in the center-of-mass frame will be of the order of the polarization potential extent. For most studied systems this will happen at temperatures in the $\mu$K regime or below. Even though this field is rapidly evolving and has promising prospects, experimental realizations are still in their nursery phase. More specifically, atom-ion interaction energy is currently limited to the mK energy scale \cite{Harter2012} and above.

In this paper, we concentrate on the aspects of elastic atom-ion collisions that currently prevail the field and limit it from entering the quantum regime. In the remaining introductory section we give an extended review on the out-of-equilibrium dynamics of elastic atom-ion collisions in Paul traps (Sec. \ref{subsec:Dynamics}). We present the results of a molecular-dynamics simulation (Sec. \ref{subsec:MD}) that captures this dynamics in the form of a power-law energy distribution of the ion's energy. This simulation treats collisions as instantaneous. Even though this simplification captures the essence of the dynamics, it misses an important aspect of atom-ion interaction in which the collisions energy is determined by the re-action of the ion to the force the atom exert on it during collision \cite{Cetina2012}. In Sec. \ref{subsec:Cetina}, we present a simulation which includes this effect and thus gives a full description of the classical elastic atom-ion collisions. Later (Sec. \ref{sec:Experiment}), we describe our hybrid experimental apparatus in which we overlap a ground-state cooled $^{88}$Sr$^+$ ion with $\mu$K cold $^{87}$Rb atoms. In Sec. \ref{sec:EMM} we discuss excess micromotion (EMM) in Paul traps. We present our method of detection and minimization of EMM that allowed us to reduce the EMM energy scale to the $\mu$K regime even for a linear ion crystal of several ions. Last and not least, in Sec. \ref{Sec:Thermometry} we present various methods for measuring the ion's energy distribution used to characterize the elastic collisions dynamics.

\subsection{\label{subsec:Dynamics}Dynamics of atom-ion collisions in a Paul trap}
In thermodynamics, two isolated systems in contact with each other will thermalize and reach the same temperature. This fundamental concept is the basis for sympathetic cooling, in which the temperature of a small system is controlled by bringing it into contact with a larger system (the bath). Sympathetic cooling of ions (small system) with atoms (the bath) can be used for cooling of dark ions, anions and molecular ions, for which the electronic structure needed for laser cooling is lacking. In the case of molecular ions, sympathetic cooling can also be used to cool rovibrational modes \cite{Tong2010,Hansen2011,Hudson2016}. Sympathetic cooling also provides a quicker and more robust way for cooling simultaneously the large number of modes of extended ion crystals for the purpose of quantum computation and simulation.

This simple picture of sympathetic cooling assumes that in each atom-ion collision energy is transferred from the hot ion to the colliding atom which is then usually lost from the shallow atoms' trap. After several collisions, this process should lead to thermalization of the ion with the atomic bath temperature. This naive picture fails in Paul-trap based atom-ion systems due to the following reasons:
\begin{itemize}
    \item Because of the time-dependent oscillating rf fields that create the Paul trap, part of the ion motion is constantly driven. In this case energy can flow in and out of the system via collisions which interrupt the driven motion of the ion with respect to the rf phase.
    \item The amplitude of the oscillating rf fields 
    grows linearly with increasing distance from the trap center. This means that the maximal energy gain in atom-ion collision increases with increasing the ion's amplitude. This results in a multiplicative random process \cite{DeVoe2009,Chen2014,Rouse2017} which results in 
    a  power-law instead of a Maxwell-Boltzmann (exponential) tail of high energies.
    \item At least three energy scales are involved in determining the ion's energy distribution. Together with the temperature of the atomic bath also the excess-micromotion energy \cite{Zipkes2011} and the re-action of the ion to the polarization potential \cite{Cetina2012} determine the steady-state energy distribution of the ion. In our system, the most dominant energy scale is the latter one.
\end{itemize}
In the following, we explain each of the above points in more detail.

To understand why a collision with a zero-temperature atom can lead to heating of the ion, it is instructive to analyze the ion trajectory in phase-space \cite{Schwarz2008}. The trajectory of an ion in an ideal Paul trap (i.e. without trap imperfections which are discussed in Sec. \ref{sec:EMM}) is given by:
\begin{equation}\label{eq:iontraj}
\begin{split}
    x_j\left(t\right)\approx&u_j\cos(\omega_{j}t+\phi_j)\left(1+\frac{q_j}{2} \cos(\Orf t)\right),\\
	v_j\left(t\right)\approx& -u_j\omega_j \sin\left(\omega_j t+\phi_j\right)\\ &-u_j\cos\left(\omega_j t+\phi_j\right) \Orf\frac{q_j}{2} \sin\left(\Orf t \right).
\end{split}
\end{equation}
This trajectory is a solution of the Mathieu equation which governs the ion's dynamics \cite{Paul1990}. Here, $\Orf$ is the frequency of the trap's oscillating rf fields, $\omega_j\approx\Orf/2\sqrt{a_j+q_j^2/2}$ are the resulting trap's secular (harmonic) frequencies. $|a_j|,q_j^2\ll1$ are the Mathieu trap-parameters which quantify the magnitude of the static ($a$) and rf ($q$) quadrupole electric fields of the trap. $u_j$ and $\phi_j$ are the amplitude and phase of the ion in its three uncoupled modes $j=x,y,z$. Eq. \ref{eq:iontraj} is a first-order approximation in the Mathieu trap-parameters. The ion's motion is composed of two terms, a slow oscillating motion at the trap harmonic frequency and fast oscillating driven motion at the trap's rf frequency which is termed micromotion. 

\begin{figure}
  \centering
  \resizebox*{8cm}{!}{\includegraphics[width=\textwidth, trim=4.1cm 7cm 4.9cm 7.9cm, clip]{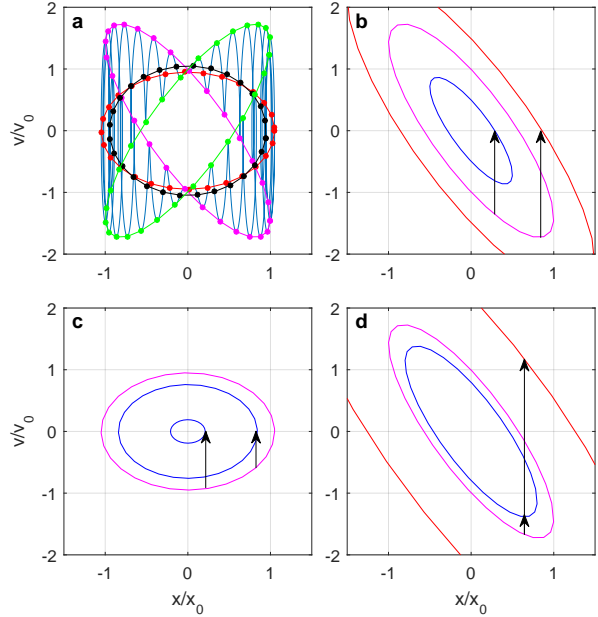}}
  \caption[Phase-space]{\textbf{Phase-space trajectory of a single ion mode in a Paul trap and the effect of hard-sphere collision.} a) In blue line we show the ion's trajectory calculated up to second-order in the trap's Mathieu parameters. Position and velocity are normalized by amplitudes, $x_0$ and $v_0=\omega x_0$. Colored dots mark the position of the ion for different rf phases ($\pi/2$ - magenta, $\pi$ - black, $3\pi/2$ - green and $2\pi$ - red). Other color lines (non-blue) are guides to the eye. b) Ion's trajectory sampled at rf phase of $\pi/2$. We show in magenta the trajectory before a collision. A collision with an equal-mass atom instantaneously removes all the kinetic energy of the ion and results in cooling (blue) or heating (red) depending on the ion distance from the trap's center. c) 
  A collision at rf phase of $2\pi$ cools the ion regardless of the ion's position in the trap (blue). d) A collision with a light atom changes the ion's energy slightly (blue) while a collision with a heavy atom can lead to significant heating (red). The phase-space picture was adopted from \cite{Schwarz2008}.}\label{Fig:PhaseSpace}
\end{figure}

The phase-space diagram of a single mode of an ion in a Paul trap is depicted in Fig \ref{Fig:PhaseSpace}a. At first glance, it looks very complicated. However, there is a simple intuitive explanation for the ion's trajectory. The ion circles on an ellipse in phase-space in its secular frequency, $\omega$, just as in the case of a simple harmonic oscillator. However, the ellipse axis wobbles at the Paul-trap's rf frequency, $\Orf\gg\omega$ due to micromotion. The ellipse also stretches and shrinks during an rf cycle. For rf phases, $\Orf t=\pi/2+\pi n$, where $n$ is an integer, the ellipse inclination is maximal (green and magenta in Fig. \ref{Fig:PhaseSpace}a) while for rf phases $\Orf=\pi+\pi n$ the ellipse inclination is zero (red and black in Fig. \ref{Fig:PhaseSpace}a) as in the harmonic case. 

For simplicity, we consider a head-on collision with a zero-temperature atom. We further simplify our discussion by assuming that the collision is instantaneous such that the ion's position remains unchanged during the collision and neglect the effect of the polarization potential attraction between the atom and the ion (this effect will be included below in Sec \ref{subsec:Cetina}). Following a collision, the ion's velocity, $v$, changes according to $v^{'}=v(m_i-m_a)/(m_i+m_a)$ from energy and momentum conservation. For heavy atoms, $m_a\gg m_i$, the velocity magnitude remains almost unchanged and collision reverses the velocity direction as in the case of hitting a wall. For light atoms, $m_a\ll m_i$, the change in velocity is small, similar to the effect of friction. Collision with equal mass atom, $m_a=m_i$, brings the ion into a complete stop. In Fig. \ref{Fig:PhaseSpace}b we illustrate how a collision with a zero-temperature atom of equal mass leads to heating if the collision occurs far from the trap's center and if the rf phase is close to $\Orf\approx\pi/2+\pi n$. For rf phase close to $\Orf\approx\pi+\pi n$, collisions will only lead to cooling as in the harmonic case (Fig. \ref{Fig:PhaseSpace}c). The heating effect is amplified for heavy atoms, $m_a\gg m_i$, due to the large momentum transfer while for light atoms $m_a\ll m_i$ heating is minimal (Fig. \ref{Fig:PhaseSpace}d).

The interplay between micromotion heating and sympathetic cooling has been studied for a long time. In 1968 Major and Demhelt \cite{Dehmelt1968} have shown that in Paul traps, sympathetic cooling is most efficient for light atoms and heavy ions (m$_i\gg$m$_a$), in contrast to ordinary sympathetic cooling which is most efficient for equal particle masses (m$_i=$m$_a$). In fact, in Paul traps, equal masses (in the Demhelt treatment) was shown to give zero cooling or heating. For light ions colliding with heavy atoms (m$_i\ll$m$_a$) it was predicted that an exponential heating of the ion will occur.

It took almost 50 years until DeVoe \cite{DeVoe2009} pointed out that the system dynamics need to be described by more than the effect of a single collision. He showed that the Major and Demhelt result of cooling or heating rate is not complete. A series of collisions far from the trap's center can heat the ion considerably above the collision energy scale (e.g atomic bath temperature). This effect includes a positive feedback in the sense that each collision that heats the ion leads to a larger trajectory and potentially larger heating in the following collision. However, the ion can always cool back very efficiently if a collision happens close to the center of the trap. This mechanism can be described using a multiplicative random process \cite{Sornette1997,Redner1990,Biro2005} which results in a power-law distribution of the ion's energy, a manifestation of non-equilibrium dynamics \cite{Stanley1988,Lutz2004}. More recently, Chen et. al. \cite{Chen2014} derived an analytical formula for the energy distribution power-law in a linear Paul with weak axial confinement, $P(E)\sim E^{-3.34 m_i/m_a+0.34}$. The weak axial confinement is defined as $q^2/a\ggg1$ where $q$ and $a$ are the Paul trap Mathieu parameters. This analytic formula agrees well with a molecular-dynamics simulation \cite{Pascal2016} performed in the same trapping regime. Additional works \cite{Rouse2015,Goodman2012} studied the same problem using numerical methods. Finally, Rouse et. al. \cite{Rouse2017} used super-statistics formalism to show how power-law statistics emerges in atom-ion systems.

In the last decade a new line of atom-ion experiments have developed, using ultracold atomic clouds \cite{Zipkes2010,Schmid2010} at $\mu$K temperatures. Initially, it was hoped that this will lead to the ground-state cooling of the ion due to the extremely low temperature of the bath. However, it was soon realized that a new energy scale emerged in the system, namely excess-micromotion (EMM). EMM is driven by non-zero rf electric fields at the Paul trap's center which results from mechanical and electrical imperfections, and can be minimized by various compensation techniques (see Sec. \ref{sec:EMM}). Zipkes et. al. \cite{Zipkes2011} performed a molecular dynamics simulation of an ion in the presence of EMM colliding with zero-temperature atoms which resulted in a power-law energy distribution. Their simulation was not limited to the weak axial trapping regime. They showed that for tight axial confinement, $q^2/a<50$, which is a pre-requisite for coherent ion manipulation in the Lamb-Dicke regime, the distribution power-law can change significantly.

The increasing understanding of atom-ion dynamics led to an interesting and fundamental question: What will be the energy distribution of an ion colliding with zero temperature atoms and no EMM? Without any energy source it seems that the ion must also go to zero energy. This naive picture does not take into account the mechanical effect of the atom-ion polarization potential on the ion position during collision. Cetina et. al. \cite{Cetina2012} performed a molecular-dynamics simulation of a zero temperature ion placed in the center of the trap colliding with zero temperature atom without EMM. In their simulation, they included the asymptotic atom-ion polarization potential, $V(r)=-C_4/2r^4$, in the numerical evaluation of the ion's and atom's trajectories. Here, $r$ is the atom-ion separation and $C_4$, which is proportional to the atomic polarizability, characterize the interaction strength. They showed that the ion is displaced from the trap center to a region of finite rf fields by the force the atom exerts on it during the collision. This mechanism introduces a new energy scale \cite{Cetina2012},
\begin{equation}\label{Eq:CetinaEnergy}
    \textrm{W}_0=2\left(\frac{\mu^5\omega^4\textrm{C}_4}{m_i^3 q^2}\right)^{1/3},
\end{equation} 
which depends on the atom and ion masses and also on trap parameters. Here, $\mu=m_a m_i/(m_a+m_i)$ is the reduced mass and $\omega$ is the trap's harmonic frequency. In our system, this energy scale is the most dominant one. We were able to verify this energy scale experimentally and also to measure the energy distribution power-law \cite{Meir2016}.

\subsection{\label{subsec:MD}Molecular dynamics simulations}
The non-equilibrium dynamics and the emergence of power-law energy distribution were captured by an analytic calculation \cite{Chen2014} and a molecular-dynamics simulation \cite{DeVoe2009,Zipkes2011,Pascal2016}. In these works, the effect of the polarization potential \cite{Cetina2012} was neglected under the assumption that the energy scale imposed by the atoms temperature or the EMM energy is dominating. We perform a similar molecular dynamics simulation which utilizes the same assumptions and concepts. Our simulation agrees with the results of these previous works. In the following, we will briefly explain the concepts behind our simulation which are similar to those in \cite{Zipkes2011} and discuss the main results pertaining ion dynamics.

We start the simulation by sampling the ion's initial total energy, $E_j$, from a given energy distribution (e.g. Maxwell-Boltzmann in the case of initially Doppler cooled ion). Here, $j=x,y,z$ refer to each of the ion's modes. We then calculate the three independent ion's amplitudes, $u_j=\sqrt{2 E_j/m_i\omega_j^2}$. Here, $\omega_j$'s are the secular mode frequencies. We sample the motion initial phases, $\phi_j$, from a uniform random distribution. 
The initial conditions determine a transient dynamics, however, if only the steady-state properties are of interest, the ion's initial state is irrelevant.

We calculate the motion of the ion between collisions using the analytic solution of the Mathieu equations in the presence of EMM: 
\begin{equation}\label{Eq:ionTrajEMM}
\begin{split}
    x_j\left(t_c\right)\approx&\left(u^{dc}_j+u_j\cos(\omega_{j}t_c+\phi_j)\right)\left(1+\frac{q_j}{2} \cos(\Orf t_c)\right),\\
	v_j\left(t_c\right)\approx& -u_j\omega_j \sin\left(\omega_j t_c+\phi_j\right)\\ &-\left(u_j\cos\left(\omega_j t_c+\phi_j\right)+u_j^{dc}\right) \Orf\frac{q_j}{2} \sin\left(\Orf t_c \right).
\end{split}
\end{equation}
Here, $u_j^{dc}$ is the ion's displacement from the trap center resulting in EMM which is in-phase with the trap rf drive \cite{Berkeland1998}. In Eq. \ref{Eq:ionTrajEMM} we neglect EMM which is quadrature with the trap drive since it is usually much smaller and plays the same role in collisions. We give a detailed theoretical treatment of EMM and explain the difference between in-phase and quadrature EMM in Sec. \ref{sec:EMM}. Since for the $r^{-4}$ atom-ion potential the rate of Langevin (spiralling) collisions is energy independent, we randomly sample the time between consecutive collisions, $t_c$, from an exponential distribution to render a constant collisions rate. We also include the effect of the atomic cloud finite-size by biasing the collision probability with the local atomic density at any instantaneous ion location \cite{Zipkes2011}. We sample the atomic energy from a Maxwell-Boltzmann distribution where the atoms velocity vector is randomly and isotropically chosen. 

Since collisions are modeled as instantaneous, the position of the ion remains unchanged following a collision, $\textbf{x}^{'}=\textbf{x}$, which implies:
\begin{equation}\label{Eq:Inst}
	u_j\cos(\omega t_c + \phi_j)=u_j^{'}\cos(\omega t_c + \phi_j^{'}).
\end{equation}
Here, $u^{'}_j$'s and $\phi^{'}_j$'s are the new ion's amplitudes and phases which are to be determined. We determine the velocity of the ion immediately after the collision using energy and momentum conservation,
\begin{equation}\label{Eq:Collision}
	\textbf{v}^{'}=\left(1-\beta\right)\left(\textbf{v}-\textbf{v}_a\right)+\beta \Re \left(\textbf{v}-\textbf{v}_a\right) + \textbf{v}_a.
\end{equation}
Here, $\textbf{v}^{'}$ and $\textbf{v}$ are the ion's velocities after and before the collision, $\textbf{v}_a$ is the atom's velocity before the collision, $\beta=m_a/(m_a+m_i)$ and $\Re$ is a rotation matrix which defines the scattering angles and it is randomly and isotropically sampled with each collision, a consequence of the Langevin model \cite{Langevin1905} which assumes that only the large angle scattering events (Langevin collisions) determines the ion's energy distribution \cite{Zipkes2011}.

We use Eq. \ref{Eq:ionTrajEMM} and Eq. \ref{Eq:Inst} to determine the secular part of the ion's velocity after a collision,
\begin{equation}\label{Eq:vprimesec}
\begin{split}
	v^{'}_{j_{|{sec}}}&\equiv
	-u_j^{'}\omega_j \sin(\omega_j t_c+\phi_j^{'})\\
	&=v^{'}_j + \left(u_j\cos\left(\omega_j t_c+\phi_j\right)+u_j^{dc}\right) \Orf\frac{q_j}{2} \sin\left(\Orf t_c \right).\\
\end{split}
\end{equation}
From Eq. \ref{Eq:Inst} and Eq. \ref{Eq:vprimesec} we can extract the ion's new amplitudes and phases and continue to the next collision.

\begin{figure}
    \centering
    \resizebox*{9cm}{!}{\includegraphics[width=\textwidth, trim=3.7cm 8.5cm 4.5cm 9cm, clip]{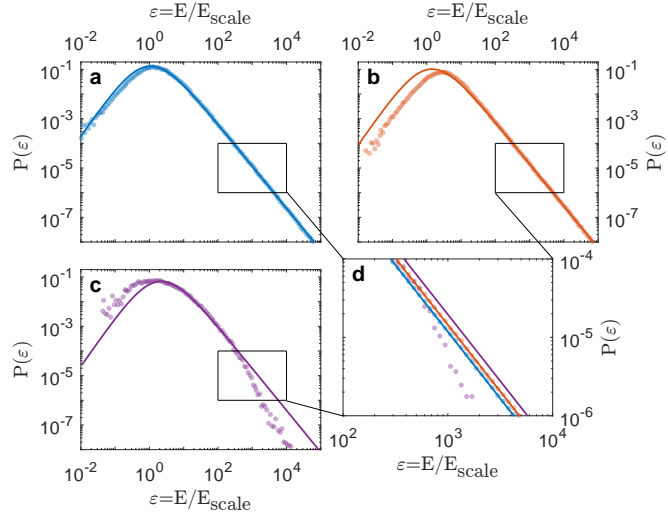}} 
    \caption[Molecular dynamics simulation]{\textbf{Molecular dynamics simulation of the ion's energy distribution for different energy sources.} Energy histogram where the energy scale is dominated by EMM energy (inset a - blue) or the temperature of the atomic cloud (inset b - red). We bin $10^7$ simulation realizations into logarithmically-spaced bins. We normalize the number of events in each bin by the bin size and total number of realizations to generate a probability distribution. Solid line is a a Tsallis distribution (Eq. \ref{Eq:Tsallis}) where we fitted only the power-law tail of the histogram. c) Energy histogram where the energy scale is dominated by the reaction of the polarization potential on the ion's position during a collision. Here, we performed only 50,000 simulation realizations. Solid line is a Tsallis distribution with $n=3.7$. The deviation of the energy histogram from the power-law tail above 300 mK is due to numerical simulation error. d) An enlarged view of the power-law tail part of the distributions. }\label{Fig:HSC simulation}
\end{figure}

In the following we summarize the main insights extracted from this kind of simulations:
\begin{itemize}
    \item The ion's energy distribution gradually develops a power-law tail, $P\left(E\right)\propto E^{2-n}$, that is determined by the system parameters, i.e., atom-ion mass ratio and the trap's Mathieu parameters. However, it is independent of the energy source, e.g., EMM energy or atoms temperature.
    \item The low energy part of the energy distribution depends on the specific energy source. When EMM energy dominates it fits nicely to a Tsallis distribution (Eq. \ref{Eq:Tsallis}) whereas other energy sources require different distributions (Fig. \ref{Fig:HSC simulation}).
    \item For power-law parameter $n\leq4$ the mean of the energy distribution diverges. For $n\leq3$ the energy distribution is not normalizable. 
    \item For stable atom-ion mixtures, the system reaches steady-state and is ergodic. 
    \item Finite atomic cloud size introduces an energy cut-off which truncates the energy distribution power-law tail \cite{Pascal2016,Dutta2017}.
\end{itemize}

In Fig. \ref{Fig:HSC simulation}a-b we show the results of such a simulation. We compare the energy distribution of the ion in steady-state for two different energy sources. We normalize the ion's energy to the energy source scale, $\varepsilon=E_\textrm{ion}/E_\textrm{scale}$. In one simulation (inset a, blue) we set the ion with a finite EMM amplitude, and we set the atoms temperature to zero. In the second simulation (inset b, red) we set the EMM amplitude to zero while we choose the atoms energy according to a Maxwell-Boltzemann distribution. For atoms temperature, the energy scale is $E_\textrm{atom}=\kB T_{a}$ where $\kB$ is the Boltzmann constant and $T_{a}$ is the atoms temperature. For EMM, the energy scale is the average EMM kinetic energy, 
\begin{equation}\label{Eq:EMMscale}
	E_\textrm{EMM}=m_{i}\left(|\mathbf{u}_\textrm{EMM}|\Orf\right)^2/4.
\end{equation}
In the case of only in-phase EMM (Eq. \ref{Eq:ionTrajEMM}), $|\mathbf{u}_\textrm{EMM}|=\sqrt{\sum_j\left(u_j^{dc}q_j/2\right)^2}$. We use the same trap parameters ($\omega/2\pi$=(0.82, 1.28, 0.58) MHz, $\Orf/2\pi$=26.51 MHz) and masses ($^{88}$Sr$^+$ ion and $^{87}$Rb atom) for both simulations. We use an infinite, homogeneous cloud to avoid finite cloud size effects.

In both simulations, we see a very clear power-law tail in the ion's energy distribution of the same magnitude for both energy sources. This can be understood since in the high energy regime of the distribution tail, inherent micromotion is much larger than $E_\textrm{scale}$ such that the power-law is determined solely by the trap's parameters and the mass ratio. The power-law becomes more prominent ($n$ decreases) as the atom-ion mass ratio increases \cite{Chen2014,Pascal2016}. For Paul trap with weak dc confinement, $q^2_j\ggg|a_j|$, the power-law magnitude, $n$, is independent of the exact value  of the trap Mathieu parameters and can be approximated by an analytic expression, $n\approx3.34(m_i/m_a-1)+5$ \cite{Chen2014}. In this regime, increasing the q-parameter increases both inherent micromotion and the radial confinement. These two effects have opposite effect on ion heating and exactly cancel. Linear Paul traps that utilize strong dc fields for axial confinement, and also for removing the radial frequency degeneracy, break this cancellation. The resulting de-confining dc fields in the radial direction  allow the ion to reach farther into regions with larger inherent micromotion. This leads to increased heating and decreases $n$. Thus in general, the power-law depends also on the trap Mathieu parameters \cite{Zipkes2011}.
\footnote{Different authors use different notion for the power-law of the energy distribution. Here it is defined as $P\left(E\right)\propto E^{2-n}$. In Chen et. al. \cite{Chen2014} it is defined as $P\left(E\right)\propto E^{-(\nu+1)}$. In Holtkemeier et. al. \cite{Pascal2016} it is defined as $P\left(E\right)\propto E^\kappa$. In Zipkes et. al. \cite{Zipkes2011} it is defined as $P\left(E\right)\propto E^{\alpha-1}$ (see remark in SM of \cite{Pascal2016}). Thus the following relation should hold when comparing between these works: $2-n=-(\nu+1)=\kappa=\alpha-1$.}

We fitted the simulated energy distribution tail to the high energy asymptote of the Tsallis energy distribution \cite{Meir2016},
\begin{equation}\label{Eq:Tsallis}
    P\left(\varepsilon\right)d\varepsilon=\frac{(n-3)(n-2)(n-1)}{2n^3}\frac{\left(\varepsilon/a\right)^2}{\left(1+\frac{\varepsilon/a}{n}\right)^n} d\varepsilon/a.
\end{equation}
As seen in Fig. \ref{Fig:HSC simulation}d, both simulations show the same distribution power-law, $n=3.7$. We analytically continued the Tsallis energy distribution to the low energy part of the distribution (solid lines in Fig. \ref{Fig:HSC simulation}). While the Tsallis function fits nicely to the entire distribution in the case of energy scale dominated by EMM (Fig. \ref{Fig:HSC simulation}a, solid blue line), it misses the bulk part when the energy source is due to atoms temperature (Fig. \ref{Fig:HSC simulation}b, solid red line). In contrast with $n$, the bulk part of the energy distribution depends on the source of energy behind $E_\textrm{scale}$.
\footnote{The low energy part of the energy distribution is dominated by the density of states. Since in our simulation, we take into account the total energy (kinetic and potential), the density of states is quadratic $P\left(E \ll   E_\textrm{scale}\right)\propto E^2$. In other work \cite{Pascal2016} only the kinetic energy was considered such that the density of states was $P\left(E \ll E_\textrm{scale}\right)\propto E^{3/2}$.}

\subsection{\label{subsec:Cetina}Ion reaction to the polarization potential}
The hard-sphere simulations of the previous section capture the dynamics of the energy scales imposed by EMM or atoms temperature. However, these simulations do not capture the dynamics related to the reaction of the polarization force on the ion's position during a collision \cite{Cetina2012}. In our system, due to careful compensation of EMM and the low atomic temperature, this is the energy scale that dominates the ion's energy distribution.

Previous numerical works \cite{Cetina2012} only looked at the effect of a single collision between a zero energy atom and a zero energy ion. Here, we perform a molecular dynamics simulation of the ion dynamics during consecutive collisions with atoms including the effect of the atom-ion polarization potential. This simulation is computationally much more demanding than the hard-sphere simulation since it requires the integration of the ion's trajectory to determine the collision outcome.

A brief description of this simulation was already given in the supplemental material of \cite{Meir2016}. Here, we repeat this description with minor changes. A detailed analysis of the simulation with respect to trap parameters and atom-ion mass ratio will be given elsewhere \cite{Pinkas2017}.

In this molecular dynamics simulation we numerically integrate the interacting atom-ion equations of motion,
\begin{equation}
\begin{split}
    \ddot{x}^{atom}_j=&2C_4\hat{x}_j\cdot\hat{\mathbf{r}}/m_a r^5\\
    \ddot{x}^{ion}_j=&(a_j+2q_j\cos(\Orf t))x_j\Orf^2/4-2C_4\hat{x}_j\cdot\hat{\mathbf{r}}/m_i r^5.
\end{split}
\end{equation}
Here, we neglect the atomic-trap potential while the ion is subject to the rf and dc fields of the Paul trap. The atom and ion are also attracted by the polarization force which pulls them towards each other ($r_j=x^{ion}_j-x^{atom}_j$). In this simulation we did not include EMM. However, it can be simply introduced by adding a homogeneous rf field to the ion's equations of motion.

We start the simulation by randomly sampling the ion's energy from a given thermal distribution and its phase from a uniform distribution. We define an interaction sphere centered around the ion trap into which atoms enter. We simulate the entrance of a single atom into the sphere at a random location on the sphere. We randomly pick the tangential velocity component with respect to the sphere from a normal distribution where the variance is proportional to the atomic temperature. The velocity component normal to the sphere is randomly picked from a Rayleigh distribution with the same temperature variance. This way we properly account for the rate in which a homogeneous gas of atoms enters to the sphere: $\Gamma=n_{a} \pi r_0^2  \sqrt{8 \kB T_a/\pi m_a}$. Here, $r_0$ is the radius of the sphere and $n_a$, $m_a$, $T_a$ are the atomic density, mass and temperature respectively.

We calculate the particles classic trajectories using the Runge-Kutta $4^\textrm{th}$ order method. We continue the integration of motion until the atom leaves the interaction sphere. In the case of a Langevin spiralling collision, we set the size of the contact interaction radius to 5 nm. At this distance, we calculate a deterministic billiard-ball collision after which we continue the integration of motion. As in \cite{Cetina2012}, this interaction can lead to a bound state of the polarization potential due to loss of energy as a result of the negative work of the rf field. Nevertheless, the atom and ion in this bound state keep colliding until positive work of the rf field increases the system energy leading to breaking of the bound state. Following the separation of the pair we continue the integration until the atom leaves the interaction sphere. We then repeat this process with a new atom where the ion's motion amplitude is kept between collisions.

In Fig. \ref{Fig:HSC simulation}c we show the results of such a simulation. We record the ion's energy after 300 Langevin collisions on average. We scale the ion's energy according to the energy scale introduced in \cite{Cetina2012} (Eq. \ref{Eq:CetinaEnergy}) which is 1.4 mK for our experimental parameters (same parameters as in Sec. \ref{subsec:MD}). We repeat the simulation 50,000 times to produce an energy histogram. The radius of the interaction sphere, into which atoms enter, sets a numerical length scale. When the ion's amplitude exceeds this length scale, we acquire a numerical error in the simulation. We set the radius of the sphere to 1.2 $\mu$m as a compromise between simulation run-time and numerical error. In Fig. \ref{Fig:HSC simulation}c we can see the numerical error in the deviation of the energy histogram from the expected power-law tail of $n=3.7$ above an energy of 300 mK. As can be seen in Fig. \ref{Fig:HSC simulation}d, similarly to the case where $E_\textrm{scale}$ is dominated by the temperature of atoms, the bulk part of low energies does not agree well with the Tsallis distribution function.

\section{\label{sec:Experiment}Experimental setup}
\begin{figure}
    \centering
    \resizebox*{9cm}{!}{\includegraphics[width=\textwidth, trim=3cm 4cm 2cm 1.5cm, clip]{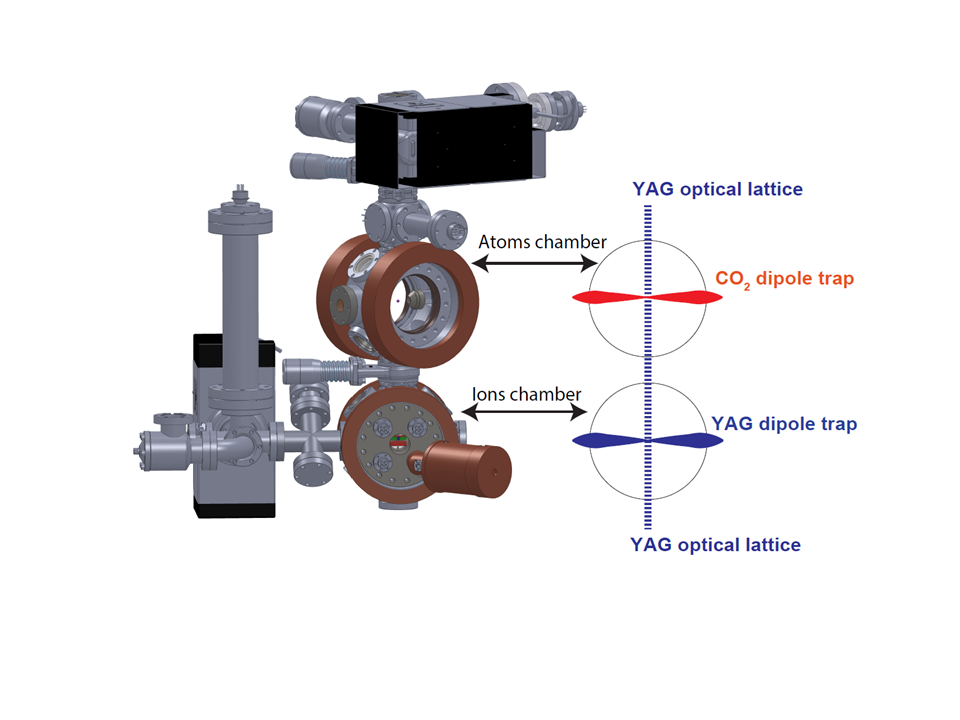}}
    \caption[Experimental system]{\textbf{Hybrid atom-ion apparatus.} Left) A SolidWorks drawing of our apparatus. Right) A sketch of the laser beams used for atomic optical trapping.}\label{Fig:Experimental system}
\end{figure}

A sketch of our experimental setup is shown in Fig. \ref{Fig:Experimental system}. The setup consists of two separate vacuum chambers connected via a thin tube to maintain a differential pressure of 1:10 between them. We use the upper chamber to collect $\sim$20,000 atoms and to cool them to a temperature of a few $\mu$K. The pressure in this chamber is $\sim10^{-10}$ Torr. We first load $\sim15$ million $^{87}$Rb atoms into a magneto-optical-trap (MOT) from a heated isotope enriched source. We then decrease the MOT beams power and increase their detuning. In this dark MOT phase the atoms density increases and temperature decreases. At this point, the atoms are loaded in the focus of an intense CO$_2$ laser beam ($\lambda$=10 $\mu$m, P=15 Watt) that is aligned with the MOT center. We then turn off the MOT beams and reduce the CO$_2$ laser power during two seconds to evaporatively cool the atoms to the $\mu$K temperature range. At the end of the evaporation, $\sim$100,000 atoms remain in the CO$_2$ trap.

To move the atoms from the upper chamber, where they are collected and cooled, to the bottom chamber, where we overlap them with the ion, we use a 1D optical lattice. The lattice consists of two counter-propagating YAG laser beams ($\lambda$=1 $\mu$m, P=5 Watt). The vertical orientation of the lattice allows us to use two collimated beams with Gaussian profiles \cite{Schmid2006}. The beams' waist is $\sim$220 $\mu$m and is overlapped with the ion trap position. The beams Rayleigh range, $z_\textrm{R}=\pi \textrm{w}_0^2/\lambda$, is comparable to the transport distance of 25 cm. Following evaporation we gradually ramp down the CO$_2$ beam power and turn on the lattice beams such that the atoms are loaded into the lattice adiabatically. We then shuttle the atoms to the bottom chamber by changing the relative frequency between the two lattice beams. The atoms' velocity is directly proportional to the relative instantaneous frequency difference, $\Delta f(t)$, between the beams, $v(t)= \lambda \Delta f(t)/2$. We use a simple triangular velocity profile during transport. We accelerate the atoms downwards for 0.1 seconds to a velocity of $\sim$0.83 m/sec and then decelerate them in 0.2 seconds back to rest. The transport itself introduces negligible atoms loss.

In principle, we can use the lattice trap to overlap the atoms with the ion. However, since the local atoms density in the lattice is modulated on a length-scale of half the optical wavelength ($\lambda/2$=0.5 $\mu$m), we choose to transfer the atoms to a cross-beam dipole trap. The atomic cloud size in this trap, $\sigma_x=\sigma_y=$5 $\mu$m, $\sigma_z=$50 $\mu$m, ensures a more homogeneous density of atoms in the vicinity of the ion. We use the lattice to move the atoms to distance of 50 $\mu$m above the ion position. We then transfer the atoms to a crossed dipole trap made of one or two lattice beams and an additional horizontal beam which is tightly focused on the ion's position ($\textrm{w}_0$=50 $\mu$m, P=1 Watt). To avoid interference between the beams, we shift the horizontal beam frequency by 240 MHz from the lattice beams frequency. We also shift the lattice beams frequency by 4 MHz from each other. Since the atomic cloud radial size is $\sim$5 $\mu$m, the atoms do not interact with the ion during the loading of the atoms to the elevated crossed-beam dipole trap 50 $\mu$m above the ion. We control the horizontal beam focus position using a piezoelectric driven mirror. To initiate atom-ion interaction, we transport the atoms downwards to overlap the ion in 5 ms. This transport method allows us to control the interaction time and local density and to perform experiments with one or less Langevin collision on average. The short transport also ensures that the ion remains in its ground-state and that it does not heat considerably due to ambient electric field noise (the heating rate of the ion is roughly 1 quanta every 20 ms). 

We chose an atomic loading and transport apparatus without the use of magnetic trapping. The reason for this choice was to avoid rf induced atomic loss from the magnetic trap due to the Paul trap rf fields.

The 1064 nm YAG beams which form the optical lattice and crossed-beam optical dipole trap shift the S$_{1/2}\rightarrow$D$_{5/2}$ quadrupole transition of the ion mostly through coupling to the D$_{5/2}\rightarrow$P$_{3/2}$ dipole transition at 1033 nm (see Fig. \ref{Fig:IonsSystem}b for the ion's energy levels). We use this shift to align the atomic dipole traps with the ion trap center using spectroscopy of the quadrupole transition with a narrow linewidth laser (see Sec. \ref{subsec:Ground-state cooling}).

\subsection{\label{subsec:Iontrap}Ion trap}
The ion Paul trap is one of the main components of the hybrid atom-ion system. Within this Paul trap ultracold atoms and ground-state cooled ions overlap and interact. We designed this Paul trap to allow for Doppler-free spectroscopy of the ion on a narrow linewidth transition and ground-state cooling of all of the ion's modes of motion. It is also designed to minimize the effects of EMM for a linear chain of ions interacting with the atoms (see Sec. \ref{sec:EMM} and Fig. \ref{Fig:EMM_Axial}).

\begin{figure}
    \centering
    \resizebox*{9cm}{!}{\includegraphics[width=\textwidth,trim={2cm 7cm 2cm 4cm},clip]{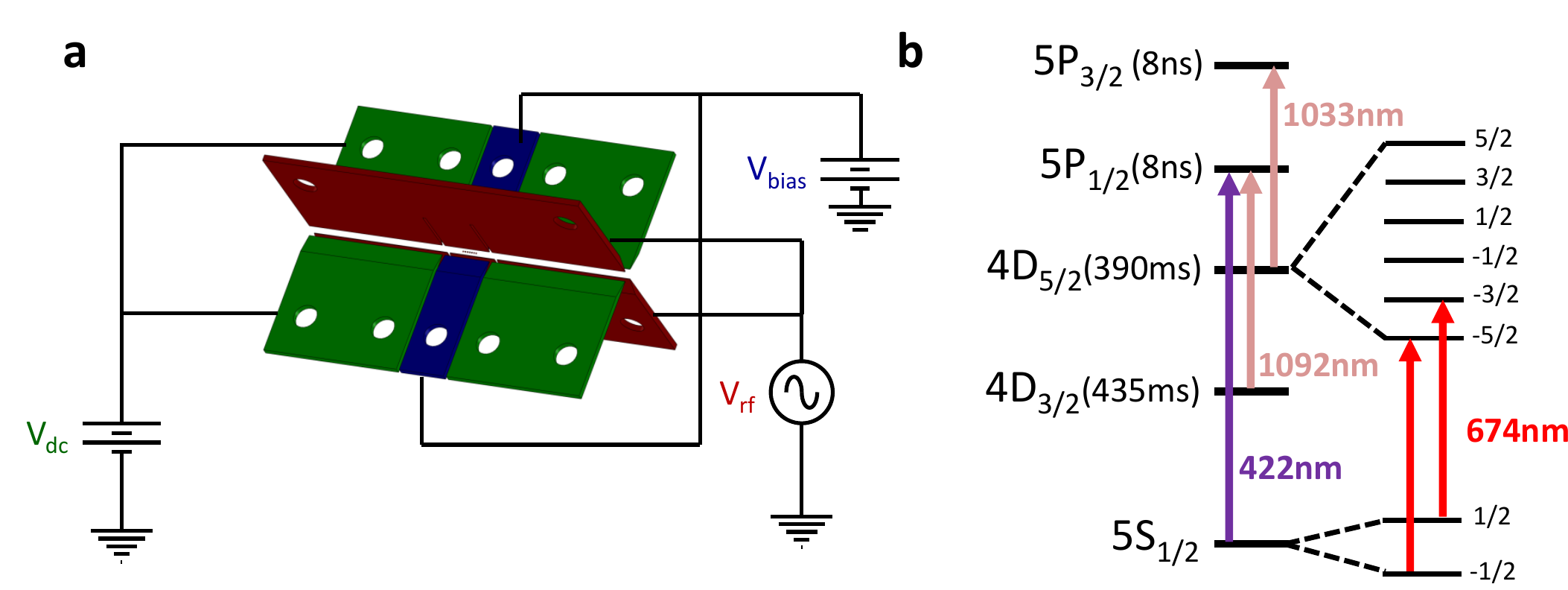}}
    \caption[Ion trap]{\textbf{a) Ion Paul trap electrodes. b) $^{88}$Sr$^+$ energy levels.}}\label{Fig:IonsSystem}
\end{figure}

The trap we use is a linear segmented Paul trap (see Fig. \ref{Fig:IonsSystem}a). The long rf-electrodes (red in Fig. \ref{Fig:IonsSystem}a) create an oscillating electric quadrupole field, $\left(q_x,q_y,q_z\right)=\left(-q,q,0\right)$, which traps the ion radially. The rf-grounded electrodes are divided into dc electrodes (green in Fig \ref{Fig:IonsSystem}a) which create a static electric quadrupole field that traps the ion axially, $\left(-a,-a,2a\right)$, and bias electrodes (blue in Fig. \ref{Fig:IonsSystem}a) which are responsible for lifting the radial mode degeneracy, $\left(-\tilde{a},\tilde{a},0\right)$. For the trap electrodes material we use 0.5 mm thick, non-magnetic, Titanium sheet which is razor sharpened on one edge. Using blades instead of rods helps maintaining the numerical aperture (NA=0.38) of the ion's imaging system (see details below). We mount the trap electrodes on a CNC-machined single mount made of a bulk of machinable Aluminium-Nitride (Shapal$^\textrm{TM}$-M) which has good thermal and rf properties.

To achieve ground-state cooling using standard techniques such as resolved-sideband cooling, we need to operate the trap in the Lamb-Dicke regime where transitions in which the ion changes its motional quantum number are suppressed. We achieve this by working at relatively high trap's mode frequencies. We use a high-Q ($\sim$500) helical resonator \cite{Siverns2012} to drive the rf electrodes at 26.51 MHz with a peak voltage ranging from 600 V to 1300 V corresponding to radial trapping frequencies of 800-1900 kHz. To achieve better stability of the trap's radial frequency, we tune the rf frequency to maximize the trap's radial frequency for a given rf power\footnote{The rf frequency that maximize the voltage on the trap electrodes and hence the ion's radial frequency differs from the rf frequency that minimizes rf reflections from the trap.}. We achieve axial trapping frequencies of 400-800 kHz by applying a dc voltage of 250-1000 V on the dc electrodes.

The relatively high voltages needed to operate the trap are a consequence of the trap's mm-scale dimensions. We use an ion-electrode distance of 0.6 mm to allow for a clear path for the atom-transport lattice collimated beams. The long rf electrodes and the segmented trap design are chosen in order to minimize the rf field gradient along the axial direction. We designed cuts in the rf electrodes that match the gaps between the dc and bias electrodes to reduce the leak of rf fields along the axial direction further. Zero rf field gradient along the axial axis implies that the trapping in this direction is purely harmonic. In the case of a linear chain of ions interacting with a cloud of atoms this will reduce unwanted micromotion induced interaction energy (see Sec. \ref{sec:EMM}).

We load ions into the trap from a home-built neutral Sr oven. We resistively heat the oven and create a beam of neutral atoms which passes through the center of the trap. We photo-ionize Sr atoms using a two-photon process with laser beams that are focused at the center of the Paul trap. The first photon is on-resonance with the $^1$S$_0\rightarrow^1$P$_1$ neutral Sr atom dipole transition at 461 nm. The second photon, at 405 nm, is tuned to a broad auto-ionization level from which the neutral atom decays into ionized Sr$^+$ and a free electron \cite{Akerman2012}. After the atom is ionized, it is immediately trapped in the deep Paul trap. Atom-ion experiments might require frequent ion loading due to inelastic processes which result in ion loss from the trap. Loading ions into the trap from a cold source \cite{Bruzewicz2016} is probably the preferred solution for atom-ion apparatuses. In our case, we implemented automatic ion loading from our hot source. We keep the Sr source current off during experiments to avoid coating of the trap's electrodes and drifts in the electric and magnetic environment of the ion \cite{Harter2014drifts}.

\subsection{\label{subsec:Doppler cooling}Doppler cooling}
We cool the ion to sub-Doppler temperatures using a 422 nm cooling beam, red-detuned from the S$_{1/2}\rightarrow$P$_{1/2}$ dipole transition together with a 1092 nm repump beam, on-resonance with the D$_{3/2}\rightarrow$P$_{3/2}$ dipole transition (see Fig. \ref{Fig:IonsSystem}b for the ion's energy levels). Since the cooling scheme involves three energy levels (eight, including the Zeeman structure), we exploit the Electromagnetically-Induced Transparency (EIT) resonance to reach temperature below the Doppler cooling limit. We scan the cooling and repump beams power and frequency across the resonance and measure the ion's steady-state temperature using coherent Rabi thermometry (see Sec. \ref{Sec:Thermometry}). At optimal laser settings, we reach a temperature of 0.3 mK. Reaching sub-Doppler temperatures improves ground-state cooling (Sec. \ref{subsec:Ground-state cooling}) efficiency.

\begin{figure}
    \centering
    \resizebox*{9cm}{!}{\includegraphics[width=\textwidth,trim={0.5cm 6cm 6cm 4cm},clip]{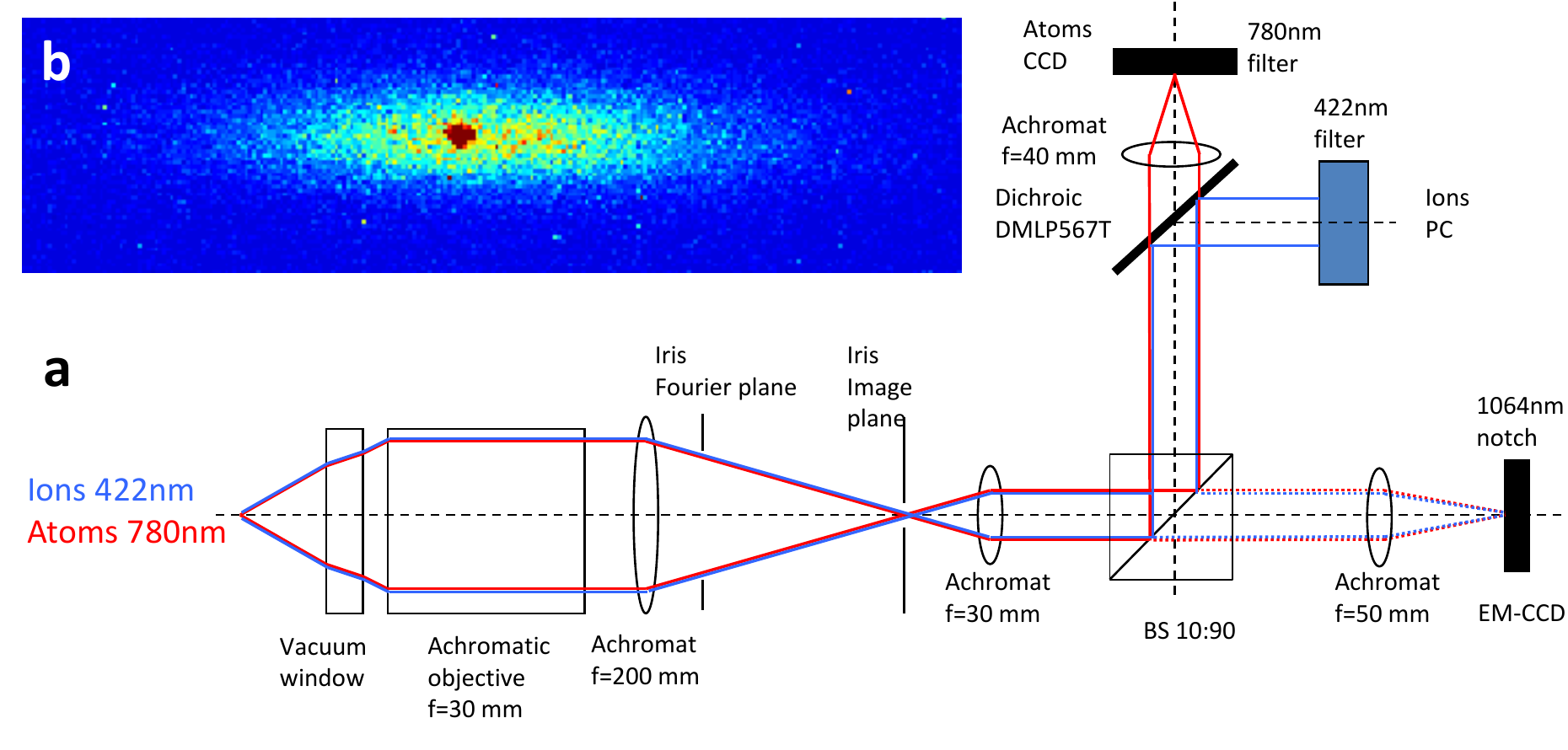}}
    \caption[Ion trap]{\textbf{a) Imaging system sketch. b) Overlapped fluorescence image of a single ion and 20,000 atoms.}}\label{Fig:ImagingSystem}
\end{figure}

We use a single bi-chromatic imaging system (see Fig. \ref{Fig:ImagingSystem}a) to collect both the 780 nm atoms' and the 422 nm ion's fluorescence. We use a special achromatic objective \cite{Alt2001} with a large focal distance (f=30 mm), a high numerical aperture (NA=0.38) and a long working distance (WD=16 mm). Ten percent of the fluorescence that is collected by the objective reaches an EM-CCD. This camera is used to acquire an overlapped fluorescence image of the atoms and the ion (Fig. \ref{Fig:ImagingSystem}b). The magnification on this EM-CCD is X10. Ninety percent of the fluorescence is deflected using a beam-splitter. The 422 nm signal is deflected again using a dichroic mirror to a photon counter (PC). We use this PC to collect fluorescence during Doppler cooling for thermometry and for state detection in coherent spectroscopy (Sec. \ref{Sec:Thermometry}). We measure the ratio between photons emitted from the ion and photons detected in the PC (collection efficiency) to be 1/(185$\pm$2) by using the ion as a single photon source. The 780 nm signal passes through the dichroic mirror and reaches a CCD. We use this CCD for the determination of the atom number and density using absorption imaging.

\subsection{\label{subsec:Ground-state cooling}Ground-state cooling}
The main tool with which we cool the ion to the ground-state, prepare it in a specific spin state and reduce its excess-micromotion (EMM), is a narrow linewidth laser (100 Hz) tuned to the optical quadrupole transition S$_{1/2}\rightarrow$D$_{5/2}$ of our Sr$^+$ ion at 674 nm (Fig. \ref{Fig:IonsSystem}b). Due to the long lifetime of the meta-stable D$_{5/2}$ state (390ms), we can coherently manipulate the ion between its electronic ground- and excited-states. The static magnetic field in our system (0.3 mT) separates the different Zeeman levels by a few MHz such that addressing individual Zeeman levels is readily achieved.

For optical pumping of the ion to the Zeeman ground-state S$_{1/2}$(m=-1/2), we tune the 674 nm laser on-resonance with the S$_{1/2}$(m=1/2)$\rightarrow$D$_{5/2}$(m=-3/2) quadrupole transition together with a repump laser on the D$_{5/2}\rightarrow$P$_{3/2}$ dipole transition at 1033 nm. After a few pulses the ion's population is pumped to the S$_{1/2}$(m=-1/2) down Zeeman ground-state with negligible population in the up Zeeman (m=1/2) state.

In order to perform ground-state cooling, we tune the 674 nm laser to the first motional red-sideband of the S$_{1/2}$(m=-1/2)$\rightarrow$D$_{5/2}$(m=-5/2) quadrupole transition together with the 1033 nm repump laser. We interlace between the different motional modes and add optical pumping pulses in between the ground-state cooling pulses. After 20 ms we can reach $\bar{n}<0.1$ in all modes.

To perform spectroscopy, we scan the 674 nm laser detuning, power and pulse time and measure the shelving probability to the D state by turning on the 422 nm cooling and 1092 nm repump beams for 1 ms after the shelving pulse. A bright signal (100 photons on average) indicates that the ion is in the S$_{1/2}$ level while a dark signal (2 photons on average) indicates that the ion was successfully shelved to the D$_{5/2}$ state.

\section{Excess-micromotion}\label{sec:EMM}
Ion's motion in a Paul trap is composed of a slow secular (harmonic) component and a fast micromotion component (Eq. \ref{Eq:ionTrajEMM}). Micromotion is usually divided into inherent-micromotion which is proportional to the harmonic oscillator amplitude and excess-micromotion (EMM) which is attributed to imperfections in the mechanical and electrical trap assembly and stray electric fields \cite{Berkeland1998}. EMM is generated by the persistence of non-vanishing rf-fields in the position of the harmonic pseudo-potential minimum. We distinguish between two EMM mechanisms which result in in-phase and quadrature oscillations with respect to the main electric quadrupole rf field phase. The compensation methods for both types of EMM are also different. While we can compensate in-phase EMM by applying dc fields, thus using the trap itself as a rf source for compensating the in-phase part of EMM, we can only compensate quadrature EMM by applying external rf fields with quadrature phase with respect to the trap rf.

We use resolved sideband spectroscopy for detecting EMM in our system. The ion's fast EMM modulation results in spectral sidebands at the rf frequency. Following Berkeland et. al. \cite{Berkeland1998} we consider a laser field in the rest frame of the ion undergoing EMM,
\begin{equation}\label{Eq:EMMlaser}
    \textbf{E}(t)=\textrm{Re}[\textbf{E}_0 e^{i\left(\textbf{k}\cdot(\textbf{u}_0+\textbf{u}_\textrm{EMM})-\omega_{l}t+\phi_{l}\right)}].
\end{equation}
Here, $|\textbf{u}_\textrm{EMM}|$ is the amplitude of EMM and $|\textbf{u}_0|$ is the amplitude of the harmonic motion. In the ion's rest frame, the laser field undergoes FM modulation, $\textbf{k}\cdot\textbf{u}_\textrm{EMM}(t)=\beta\cos(\Orf t + \delta)$, with modulation index, $\beta=\sqrt{\left(\sum_i \textrm{k}_i \textrm{u}_i^\parallel\right)^2+\left(\sum_i \textrm{k}_i \textrm{u}_i^\perp\right)^2}$, and phase shift, $\delta=\tan^{-1}\left(\frac{\sum_i \textrm{k}_i \textrm{u}_i^\perp}{\sum_i \textrm{k}_i \textrm{u}_i^\parallel}\right)-\frac{\pi}{2}$. The EMM amplitude, $\textbf{u}_\textrm{EMM}=\textbf{u}^\parallel\cos(\Orf t)+\textbf{u}^\perp\sin(\Orf t)$, is composed of in-phase micromotion, $|\textbf{u}^\parallel|$, which oscillates in-phase with the trap rf frequency and quadrature micromotion, $|\textbf{u}^\perp|$, which oscillates 90 degrees out-of-phase with respect to the rf frequency.

We can re-write Eq. \ref{Eq:EMMlaser} in terms of an infinite Bessel function series expansion, $e^{i\beta\cos(\Orf t + \delta)}=\sum_n{\textrm{J}_n(\beta)}e^{in\left(\Orf t + \delta+\pi/2\right)}$,
\begin{equation}
    \textbf{E}(t)=\textrm{Re}[\sum_{n=-\infty}^\infty{\textbf{E}_0\textrm{J}_n(\beta)}e^{i \left(\textbf{k}\cdot\textbf{u}_0 - \left(\omega_{l}-n\Orf \right)t + 
    \phi_{l} + \delta +n\pi/2 \right)}].
\end{equation}
From the above expression, we see that the spectrum acquires sidebands at, $\omega_{0}=\omega_{l}-n\Orf$, where $n$ in an integer. The amplitude of the $n$'th sideband is proportional to $\textrm{J}_n(\beta)$. Since the modulation index, $\beta$, for our residual EMM is small ($\beta<0.1$) we neglect all but the carrier and first order sidebands ($n=0,\pm1$). Since the trap's rf frequency is much larger than the quadrupole transition linewidth and the laser coupling strength (Rabi frequency), the sidebands are resolved.

Our method of detecting EMM is based on detecting the coupling of our laser to these resolved sidebands. We perform a shelving experiment on the EMM sideband and carrier and measure their relative couplings,
\begin{equation}\label{Eq:modindex}
    \frac{\Omega_1}{\Omega_0}=\frac{\textrm{J}_1(\beta)}{\textrm{J}_0(\beta)}\approx\frac{\beta}{2}.
\end{equation}
Here, $\Omega_{0,1}$ denote the carrier and sideband Rabi frequencies respectively. This method measures the EMM amplitude along the laser $\textbf{k}$-vector such that the following relation is satisfied,
\begin{equation}\label{Eq:EMM_amplitude}
	\beta = \textbf{k} \cdot \textbf{u}_\textrm{EMM}
\end{equation}

\begin{figure}
	\centering
	\resizebox*{\textwidth}{!}{\includegraphics[width=\textwidth,trim={4cm 13cm 3.5cm 8cm},clip]{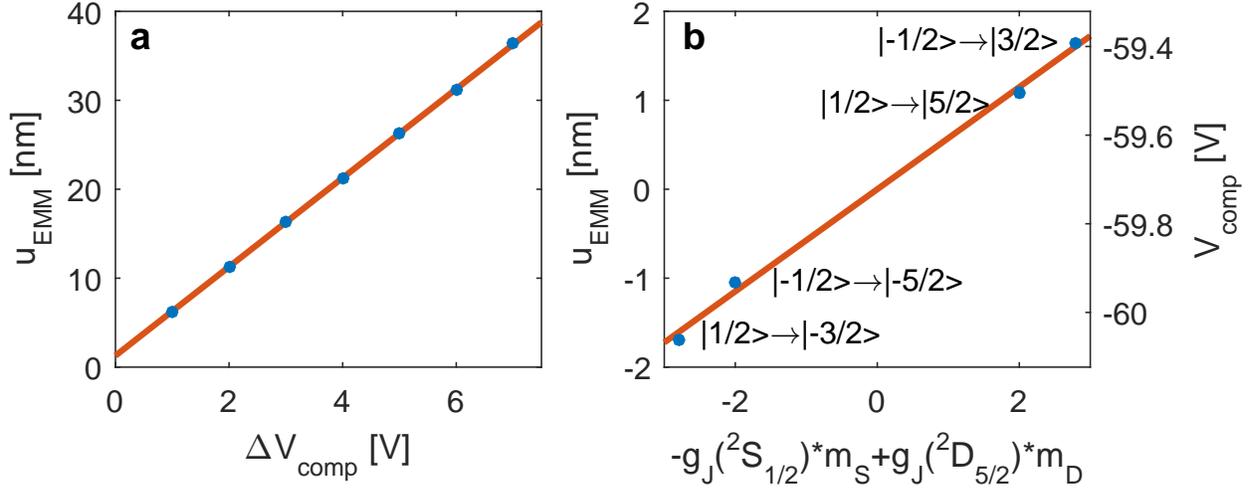}}
	\caption[EMM]{\textbf{a) EMM response.} Total EMM amplitude, $|\mathbf{u}_\textrm{EMM}|$, as a function of the difference in the compensation voltage, $\Delta\textrm{V}_\textrm{comp}$, from the compensated value. We use two orthogonal beams in the radial plane on the S$_{1/2}(-1/2)\rightarrow$D$_{5/2}(-5/2)$ transition. The linear slope is 4.98$\pm$0.04 nm/V. The EMM amplitude is not zero at the compensated point due to rf-Zeeman induced systematic shift. The error-bars are too small to show in this figure. \textbf{b) Rf-Zeeman induced systematic shift.} Absolute compensation voltage (right y-axis) needed to minimize the sideband coupling for different Zeeman sub-levels transitions. The x-axis is proportional to the linear susceptibility of the S$_{1/2}\rightarrow$D$_{5/2}$ Zeeman sub-levels. g$_\textrm{J}$ is the Land\'e g-factor. Red line is a linear fit from which we extract the correct EMM compensation value of -59.72$\pm$0.02 V. We use the EMM response calibration to translate the compensation voltage shifts to EMM amplitude (left y-axis). The error-bars are too small to show in this figure.}\label{Fig:EMM}
\end{figure}

An example of such a scan in which we extract the EMM amplitude in the radial plane is depicted in Fig. \ref{Fig:EMM}a. We measure the full EMM amplitude as a function of one of the radial dc-compensation electrodes voltage using two orthogonal beams. From the linear slope we extract the EMM response to uncompensated voltage in this electrode to be 4.98$\pm$0.04 nm/V for the specific trap parameters used in this calibration. For the other radial dc-compensation electrode the response is 4.01$\pm$0.14 nm/V. We use these calibrations later to estimate the residual EMM in our system due to compensation uncertainties and drifts.


Our method for detecting EMM is sensitive to the total amplitude of EMM resulting from both in-phase and quadrature EMM. Other methods such as parametric excitation \cite{Ibaraki2011}, ion trajectory analysis \cite{Gloger2015} and CCD optimization are sensitive only to the in-phase part of EMM since they rely on the displacement of the ion from the trap center which is generated only by stray dc fields. To detect EMM in our system we use three 674 nm laser beams with linearly independent $\textbf{k}$-vectors, each measures a different projection of the EMM amplitude. This way we analyze the total EMM amplitude in our system.

\subsection{Rf-Zeeman systematic shift}
Resolved EMM sideband spectroscopy is based on detecting the modulation of the transition frequency at the rf frequency. In the case of EMM, the Doppler shift associated with the ion's oscillatory motion induces this frequency modulation (FM). However, any other source of FM of this transition will also be detected in the same way and will be identical to the EMM signal. All these sources will introduce a systematic error to the evaluation of the EMM magnitude and the determination of the experimental settings that null EMM. 

One source of such systematic error is an oscillating magnetic field at the position of the ion, induced by rf currents running in the vicinity of the ion trap. Due to this systematic shift, the coupling strength of the EMM sidebands change for different Zeeman sub-level transitions. In Fig. \ref{Fig:EMM}b we show this effect for different S$_{1/2}$(m$_\textrm{S}$)$\rightarrow$D$_{5/2}$(m$_\textrm{D}$) Zeeman transitions. We see that for different Zeeman transitions, we require a different EMM compensation voltage to null the sideband coupling. The correct compensation value is given by extrapolating to the (non-existing) non-susceptible transition (m$_\textrm{S}$=m$_\textrm{D}$=0). In practice, we determine the EMM compensation value from the average of two measurements on two opposite-sign transitions.


The magnitude of the rf-Zeeman systematic shift depends on the Zeeman susceptibility of the transition and the oscillating magnetic field amplitude at the position of the ion. In our apparatus, the magnitude of this effect is $\beta_\textrm{Zeeman}\approx0.01$ which in our laser configuration of 45$^\circ$ translates to 1.5 nm EMM amplitude in the radial direction (see Fig. \ref{Fig:EMM}b). The energy associated with this motion (Eq. \ref{Eq:EMMscale}) is roughly 150 $\mu$K. This means that without taking into consideration this systematic shift we add this amount of energy to our system. To the best of our knowledge, there is no discussion or measurement of the rf-Zeeman systematic shift in the literature. The systematic magnitude is relatively large and can bias the uncertainties of ion-based optical clocks standards as well.

The uncertainty in the determination of the compensation voltage using this method should be taken into account in the residual EMM budget. Typically, the voltage compensation uncertainty is roughly 20 mV. This translates to EMM amplitude of 0.1 nm and energy of less than 1 $\mu$K. However, when repeating the EMM compensation processes the standard-deviation of the compensation values could reach up to 80 mV. This accounts to 0.4 nm and 12 $\mu$K. We can improve on this uncertainty by integrating more in each of the compensation scans. 

\subsection{Temperature systematic shift}
Another possible source of uncertainty is due to the finite ion temperature \cite{Keller2015}. Due to the non-linear laser-ion coupling, second-order coupling between inherent micromotion and harmonic motion sidebands overlaps the EMM sideband at the rf frequency. The relative coupling between the carrier and sideband due to this effect is given by,
\begin{equation}\label{Eq:EMMTempCpl}
	\frac{\Omega_1}{\Omega_0}=2\frac{\textrm{J}_1(ku)}{\textrm{J}_0(ku)}\frac{\textrm{J}_1(kuq/4)}{\textrm{J}_0(kuq/4)}\equiv\frac{\beta_\textrm{temp}}{2}.
\end{equation}
Here, $u$ is the harmonic amplitude (Eq. \ref{Eq:ionTrajEMM}) and $k$ is the projection of the laser k-vector in the radial plane. There are two equal contributions for the coupling from $(\Orf-\omega)+\omega$ and $(\Orf+\omega)-\omega$. Using the Lamb-Dicke parameter, $\eta=k\sqrt{\hbar/2m\omega}$, and the classical definition of the ion's total energy, $\kB\textrm{T}_\textrm{ion}=m\left(u\omega\right)^2/2$ we can rewrite Eq. \ref{Eq:EMMTempCpl} as,
\begin{equation}\label{Eq:EMMTemp}
	\beta_\textrm{temp}\left(\textrm{T}_\textrm{ion}\right)=q\eta^2 \kB \textrm{T}_\textrm{ion}/\hbar\omega.
\end{equation}
For our experimental parameters at the sub-Doppler temperature, $\beta_\textrm{temp}\approx0.001$ which is equivalent to EMM energy of roughly 1 $\mu$K. This systematic error is restricted to the radial plane since in the axial direction trapping is almost purely due to static electric fields such that the inherent micromotion sidebands are negligible.

\begin{figure}
    \centering
	\resizebox*{\textwidth}{!}{\includegraphics[width=\textwidth,trim={4cm 12cm 3.5cm 9cm},clip]{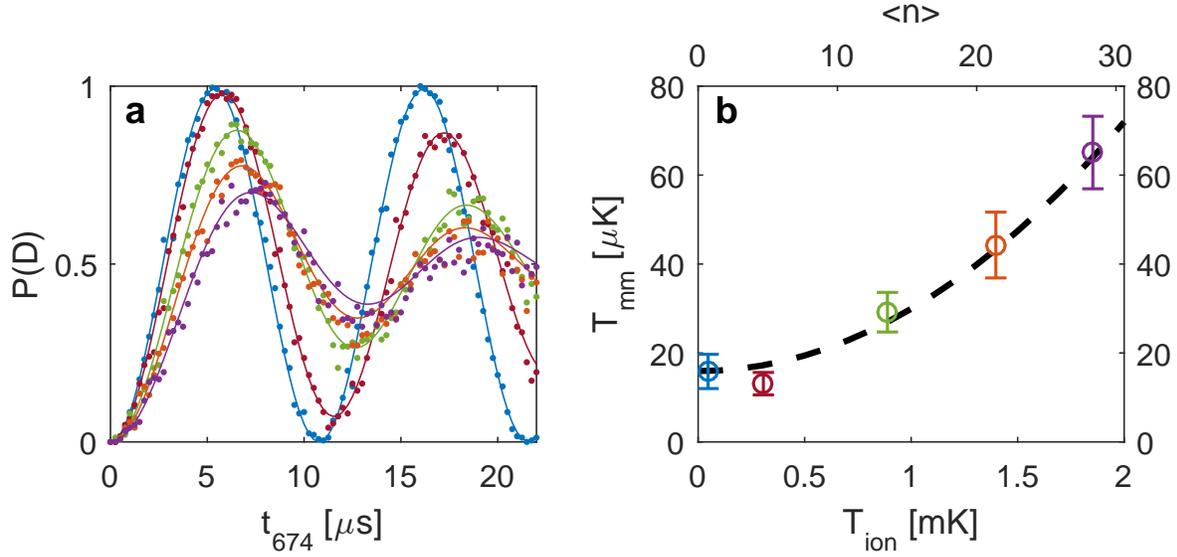}}
	\caption[Temperature induced systematic shift]{\textbf{a) Temperature calibration.} Carrier Rabi spectroscopy for different cooling parameters. In blue we use sideband cooling to cool the ion to the ground-state. In red we cool the ion using optimal sub-Doppler cooling. In green, orange and purple we change the cooling parameters to increase the ion's temperature. The solid line is a fit to a carrier Rabi flopping model (sec. \ref{Sec:Thermometry}) which assumes a thermal distribution with two fit parameters: bare Rabi coupling and temperature. \textbf{b) Temperature induced systematic shift.} Micromotion temperature (Eq. \ref{Eq:EMMscale}) as function of the ion's temperature. Color code is the same as in caption a. The dashed line is the theoretical prediction of Eq. \ref{Eq:EMMTemp} with a constant offset fit-parameter to account for uncompensated quadrature EMM. The offset value is found to be 15.9$\pm$1.2 $\mu$K. Data are for the projection of the EMM on the $\mathbf{k}$-vecotr of a single detection beam in the radial plane.}\label{Fig:EMM_Temp}
\end{figure}

We measure the temperature induced systematic error by measuring the EMM sideband coupling for ion temperatures ranging between ground-state ($<$50 $\mu$K) to 2 mK (Fig. \ref{Fig:EMM_Temp}). The temperature scaling agrees with the theoretical prediction of Eq. \ref{Eq:EMMTemp}. In contrast with the rf-Zeeman error, the error due to finite temperature does not shift the in-phase EMM compensation value \cite{Keller2015,Keller2015conf}. Instead, it only biases the uncompensated EMM amplitude estimation which is important for determining, e.g, atom-ion collision energy in hybrid systems or second-order Doppler shifts in atomic-ion optical clocks. In our system, the EMM sideband coupling for 2 mK ion results in an overestimation of the EMM energy by roughly 50 $\mu$K (red in Fig. \ref{Fig:EMM_Temp}). With sub-Doppler cooling temperatures (0.3mK) and below, this systematic error becomes negligible. 

After ground-state-cooling (blue in Fig. \ref{Fig:EMM_Temp}), we still measure weak coupling on the EMM sideband. The source of this residual coupling is likely due to drifts in the dc-compensation voltage and quadrature EMM. The minimal atom-ion collision energy we can therefore achieve is bound by this offset to $\sim30$ 
$\mu$K, considering the two radial modes. When EMM compensation is close to optimal, sideband couplings are extremely weak, on the order of kHz and below. Technical noises such as magnetic field fluctuations decohere the resulting Rabi nutation. We therefore cannot extract the sideband Rabi frequency, $\Omega_1$, from a full Rabi cycle. Instead, we measure the shelving probability for short pulse time of 25 $\mu$s. Due to extremely low shelving probability we average over 300,000 experimental realizations to get sufficient statistical significance.

\subsection{Quadrature EMM}
Compensating the in-phase part of EMM is a well known procedure and common practice in the ion trapping community. By using dc-compensation electrodes, we move the ion in the trap to a position where the rf fields generated by the trap are opposite and equal in magnitude to the EMM fields. Since the trap's rf has a definite phase, this technique only diminishes the in-phase part of EMM. Moreover, this method relies on the fact that the trap creates a gradient of rf electric fields. This is the case in the radial directions. However, in the axial direction in linear symmetric traps, there is no rf gradient. We should emphasize that this does not imply that rf fields are not present in the axial direction. Constant rf fields can penetrate into the axial axis due to assembly imperfections such as a wedge angle between the rf electrodes \cite{Dawkins2016}. These axial fields create EMM which we cannot compensate using the dc compensation electrodes. 

In the axial direction, we measure EMM using a single 674 nm laser beam aligned directly with the axis such that we are only sensitive to motion in the axial direction. Without any compensation, we measure $\beta_\textrm{axial}=0.035$. The ion oscillates with an amplitude of 3.7 nm which corresponds to a kinetic energy of 1 mK. Moving the ion along the axial direction by 10's of $\mu$m to each side, does not change the EMM amplitude. This is because our trap is symmetrically designed to minimize the rf gradient field along this direction. Using a simple driven harmonic oscillator model, we can derive the rf field amplitude in the axial direction, \begin{equation}\label{Eq:Erf}
    \textrm{E}_\textrm{rf}=m\Orf^2|\textbf{u}_\textrm{EMM}|/e,
\end{equation}
which for our EMM amplitude corresponds to 94 V/m amplitude of oscillating electric field. Here, $e$ is the electric charge of an electron.

\begin{figure}
    \centering
    \resizebox*{\textwidth}{!}{\includegraphics[width=\textwidth,trim={4cm 12cm 3.5cm 8cm},clip]{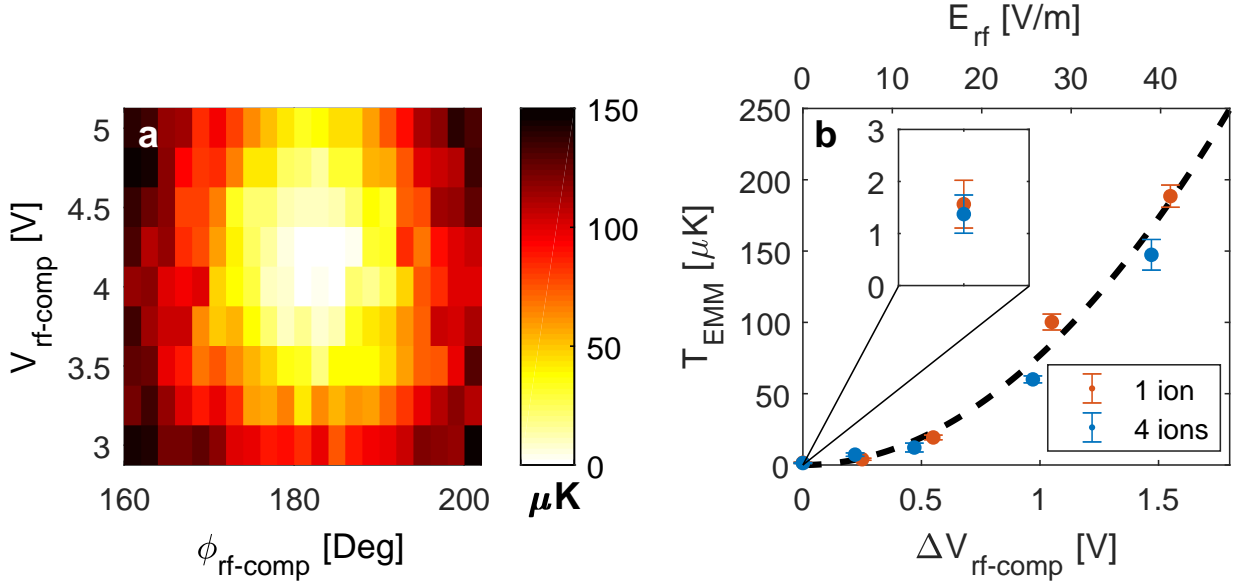}}
	\caption[EMM]{\textbf{a) Axial compensation.} 2D scan of the axial rf compensation field amplitude (y-axis) and phase (x-axis). The color bar is an estimation of the EMM kinetic energy in $\mu$K. \textbf{b) Residual Axial EMM.} A measurement of the EMM energy per ion as a function of the rf electric field at the position of the ions. For single (red) and four (blue) ions the EMM coupling per ion is the same. Dashed black line is an extrapolation of the EMM energy from the uncompensated value of 1mK. The inset shows the residual axial EMM for single (1.6$\pm$0.5 $\mu$K) and four ions (1.4$\pm$0.4 $\mu$K) after rf-compensation.}\label{Fig:EMM_Axial}
\end{figure}

To compensate for this constant field, we inject an oscillatory rf field in the axial direction. We scan the rf compensation field amplitude and its phase relative to the trap's rf (Fig. \ref{Fig:EMM_Axial}a). Since the magnitude of the non-compensated rf field in the axial direction is considerably large, we amplify the compensation rf field using an amplifier (5W) and a low-Q (Q$\sim$70) resonator. After we apply the rf compensation, we are left with almost undetectable micromotion in the axial direction. 

In Fig. \ref{Fig:EMM_Axial}b we show a measurement of the residual EMM in the axial direction after rf-compensation. We measure the shelving probability on the EMM sideband using a short (27 $\mu$s) pulse. We detect an amplitude of 0.15 nm which corresponds to 1.5 $\mu$K EMM energy when micromotion is compensated. This corresponds to a reduction of the EMM energy by three orders of magnitude by rf-compensation. 

To check the homogeneity of the electric field in the axial direction we perform the same measurement on a four-ions crystal instead of a single ion (blue data points). The crystal length was 17.5 $\mu$m \cite{James1998}. The detected EMM per ion for four ions is the same as the EMM for a single ion which indicates high degree of rf field homogeneity along the trap axis.   

We summarize the many aspects of EMM measured in our system in Table \ref{Table:EMM}.

\begin{table}
\tbl{\textbf{EMM budget.} Different sources of EMM and their respective amplitudes and corresponding temperatures (Eq. \ref{Eq:EMMscale}).\\}
{	\begin{tabular}{ | p{6.2cm} | p{1cm} | p{1cm} |}
		\hline
		Description & $\textbf{u}_\textrm{EMM}$\textsuperscript{a} [nm] & T$_\textrm{EMM}$ [$\mu$K]  \\ \hline
		Axial EMM without rf-compensation & 3.7 & 1000 \\ 
		\hline
		Residual Axial EMM for single ion & 0.15 & 1.5 \\ 
		\hline
		Residual Axial EMM for four ions & 0.15 & 1.5 \\ 
		\hline
		rf-Zeeman systemic shift & 1.5 & 150 \\ 
		\hline
		Temperature systematic @ 0.3 mK & 0.1 & 1 \\
		\hline
		Uncertainty in radial compensation value & 0.4 & 12 \\
		\hline
		Residual radial EMM after dc-compensation & 0.5 & 16 \\
		\hline
		\hline
		Total EMM in our apparatus\textsuperscript{b} & 0.7 & 32 \\
		\hline
	\end{tabular}
}
\tabnote{\textsuperscript{a}To calculate the oscillating rf electric field, E$_\textrm{rf}$, at the ion position use Eq. \ref{Eq:Erf}. For our trap parameters, E$_\textrm{rf}$ [V/m] $\approx25 |\mathbf{u}_\textrm{EMM}|$ [nm].}
\tabnote{\textsuperscript{b}We linearly sum the EMM energies contribution from the axial and the two radials, which corresponds to a quadrature sum of the fields.}
\label{Table:EMM}
\end{table}

\section{Thermometry with ions}\label{Sec:Thermometry}
\begin{figure}
    \centering
    \resizebox*{9cm}{!}{\includegraphics[width=\textwidth,trim={4cm 7cm 5cm 7cm},clip]{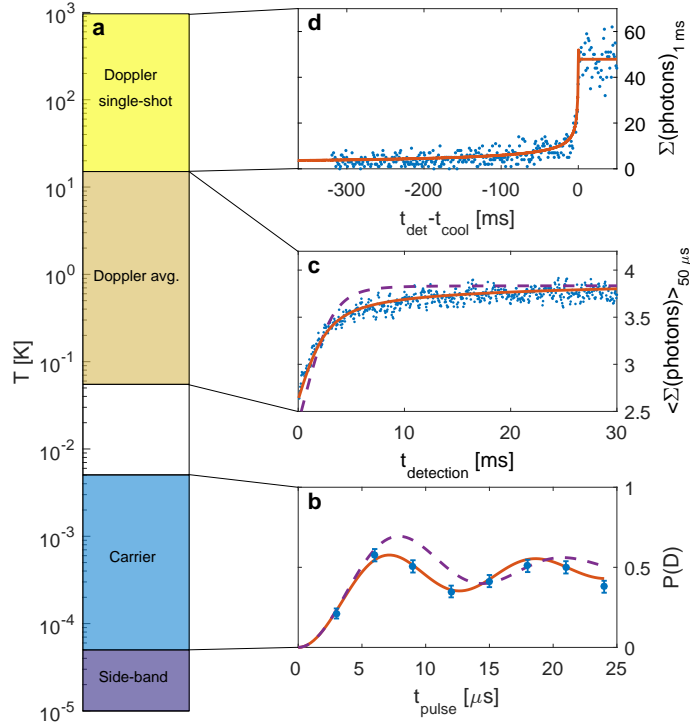}}
    \caption[Ion thermometry]{\textbf{Ion thermometry.} a) Thermometer bar which indicates the energy span of four thermometry methods used in our experiment (experimental data for sideband thermometry are not shown). b) Carrier Rabi thermometry. The blue data points are the excited state probability. Each data point is averaged over 170 repetitions. Error bars are one-sigma binomial error. Red line is a fit using a Tsallis distribution (Eq. \ref{Eq:Tsallis}) while dashed purple line is a fit using a Maxwell-Boltzmann distribution. Data are adopted from \cite{Meir2016} c) Average Doppler cooling thermometry. The blue points are the number of photons detected in intervals of 50 $\mu$s averaged over 350 experimental realizations. We add a 150 $\mu$s moving average filter to reduce data shot-noise. The red line is a result of a model where we use a Tsallis distribution for the ion's energy. Dashed purple line is the same model but with a Maxwell-Boltzmann distribution. Both models use a single fit parameter. Data are adopted from \cite{Meir2016} d) Single-shot Doppler cooling thermometry. The blue points are the number of photons detected at 1 ms intervals, in a single experimental realization. The red line is the result of a simulation with a single fit parameter (initial energy).}\label{Fig:Thermometer}
\end{figure}

In atom-ion experiments, the ion's energy can span many orders of magnitudes. Before interaction with the atoms, we prepare the ion in its ground-state where the mean harmonic oscillator occupation is $\bar{n}<0.1$. For our trap parameters, this accounts for energy in the $\mu$K range. In our system, due to the energy imparted by the atom-ion polarization force during collision (Sec. \ref{subsec:Cetina}), the ion heats in steady-state to the mK energy range. Inelastic processes such as hyperfine changing collisions or quenching from electronic excited state \cite{Ratschbacher2012} release energy ranging from 100's mK to 1000's K. To evaluate the energy involved in elastic and inelastic atom-ion collisions we used various thermometry methods each suited for a different energy regime (Fig. \ref{Fig:Thermometer}). 
In the low energy scale, we use a narrow laser (100 Hz) on an electric quadrupole transition at 674 nm (Fig. \ref{Fig:IonsSystem}b) to detect the ion's motional sidebands \cite{Leibfried2003}. This is a well known technique which can be used to characterize the ion's Fock-state distribution near the ground-state \cite{Kienzler2015}. We used this method to analyze the ion's ground-state population following cooling, measure the ion's heating rate due to electric field fluctuations and measure the angle of the different modes with respect to the laser $\textbf{k}$-vector. 

When the mean harmonic oscillator occupation becomes larger than unity, the sideband method loses its sensitivity. For this energy regime, we evaluate the ion's energy distribution by coherently driving the carrier transition (Fig. \ref{Fig:Thermometer}d). The laser induced carrier-coupling of the ground and excited electronic states depends on the harmonic oscillator occupation,
\begin{equation}
    \Omega_{\mathbf{n}\rightarrow\mathbf{n}}=\Omega_0\prod_i e^{-\eta_i^2/2} \textrm{L}_{n_i}\left(\eta_i^2\right).
\end{equation}
Here, $\textrm{L}_{n_i}(x)$ is the Laguerre polynomial of degree $n_i$, where $n_i$ is the harmonic oscillator occupation of the i-th mode, and $\eta_i$ is the Lamb-Dicke parameter of the mode. For this reason, broad energy distributions such as a power-law distribution result in fast decoherence of the sinusoidal oscillation between the two electronic states. A detailed derivation can be found in the supplemental material of \cite{Meir2016}.

In Fig. \ref{Fig:Thermometer}b we show the results of carrier thermometry performed on the ion after 15 atom-ion collisions on average. We measure the excited state population for different laser pulse times. We show that the data agrees well with a Tsallis distribution (Eq. \ref{Eq:Tsallis}, red solid line in figure) with a power-law parameter $n=4$, while it disagrees with a Maxwell-Boltzmann energy distribution (dashed purple line). Data are adopted from \cite{Meir2016}.

The carrier thermometry method can be used up to a temperature of a few mK until it also losses its sensitivity as the oscillations contrast is too small for a reasonable signal to noise ratio (we are usually dominated by projection noise). At this energy range from few mK up to 10's mK, coherent thermometry fails while non-coherent methods, which are discussed below, are still insensitive. Alternative thermometry methods such as exploiting the EIT resonance \cite{Roßnagel2015}, power-broadening a narrow linewidth transition \cite{Hendricks2008} or using ultra-fast laser pulses \cite{Johnson2015} were not explored in the context of this work.

At energies above 50 mK, Doppler shifts are large enough to reduce the resonance fluorescence signal from the dipole-allowed cooling transition at 422 nm (Fig. \ref{Fig:IonsSystem}b). A Model for the fluorescence during Doppler cooling involving only two energy levels can be found in \cite{Wesenberg2007}. In our case, we needed to modify this model to take into account the eight energy levels which participate in the cooling process of our ion and also to account for the change in the spectrum due to inherent micromotion. Details regarding our model can be found in \cite{Sikorsky2017}.

In Fig. \ref{Fig:Thermometer}c we show the results of Doppler cooling thermometry performed on the ion after 100's of atom-ion collisions. EMM was deliberately set to 150 mK to dominate all other energy scales. We record the fluorescence in 50 $\mu$s time bins for 30 ms. Since we detect $\sim$3.5 photons on average for each time-bin we needed to repeat the experiment 350 times to reduce photon shot-noise. We show that the data agrees with our model when we use a Tsallis distribution (red line in the figure) while it disagrees with Maxwell-Boltzmann energy distribution (dashed purple line). Data are adopted from \cite{Meir2016}.

At energies up to few K's the signal-to-noise from a single cooling trajectory is not enough to determine the ion's energy due to photon shot-noise. For higher energies, the fluorescence signal drops significantly below the level of Doppler cooled ion. While the ion cools down the fluorescence does not increase much until the ion reaches again the few K temperature regime where fluorescence increases in few ms (see Fig. \ref{Fig:Thermometer}b). The sharp feature in the non-linear cooling trajectory allows us to extract the ion's initial energy from the time it took the ion to cool in a single experimental realization (single-shot). By repeating the experiment with the same initial conditions, we can directly reconstruct the ion's energy distribution without the need for a theoretical model. We use this technique to directly observe sympathetic cooling and non-equilibrium dynamics in our system \cite{Meir2017neq}. 

At the high energies of the single-shot Doppler cooling technique, the model which we used in the low energy regime fails since it assumes that the excited state population is in steady-state which is not the case for high energies. Lacking a simple model for Doppler cooling, we use a brute-force approach in which we numerically solve the ion's equation of motion in the presence of a scattering force which is proportional to the excited state population. The excited state population is evaluated simultaneously by solving the 64 coupled dynamical Bloch equations. We also included in the numerical simulation the finite cooling and repump beam sizes and the trap anharmonicity. Details regarding this simulation can be found in \cite{Meir2017cooling}.

In Fig. \ref{Fig:Thermometer}d we show the results of a single Doppler cooling trajectory of an ion initialized in a classical coherent state with 170 K energy. We record the fluorescence in 1 ms time-bins for 200 ms. The red line is the fluorescence given by our simulation using a single fit parameter (initial energy).

\section{Conclusion}
In this paper, we describe an apparatus for overlapping ground-state cooled ions with ultracold atoms such that both species are in the $\mu$K energy regime before interaction. Even though we focus in this work on the single ion-atoms interaction, our EMM compensation allows us to interact with linear ion crystals of several ions in the $\mu$K energy regime. We give an extensive review on the single ion's dynamic in this system and also present numerical tools to simulate this dynamic. 

Several energy scales are involved in determining the result of an atom-ion elastic collisions. To reach the ultracold regime of atom-ion collisions, EMM energy must be reduced to the $\mu$K scale. We perform an extensive analysis of the EMM in our system. We show methods to compensate and evaluate EMM. We discuss different systematic errors which can bias the EMM compensation and increase the interaction energy. With this methods we decrease the EMM in our system to the $\mu$K scale. Careful analysis and compensation of EMM is also crucial for optical clocks based on trapped ions.

Atom-ion collisions result in a change of the ion's kinetic energy. Here, we presented three ion thermometry methods which span from $\mu$K to 1000's K energies. We use these methods to determine the ion's energy distribution after interacting with the atoms. Thermometry is also crucial tool for evaluating in-elastic and chemical processes which is an essential part in atom-ion experiments \cite{Sikorsky2017sr,Benshlomi2017}.

This work focused on the interaction between atoms and ions where the ions are trapped in a Paul trap. Up to this date, all experimental realizations of ultracold atom-ion experiments are based on Paul traps. Other approaches utilize multi-pole traps \cite{Pascal2016} and optical traps \cite{Lambrecht2016} are under development.

This work was supported by the Crown Photonics
Center, ICore-Israeli excellence center circle of light, the
Israeli Science Foundation, the U.S.-Israel Binational
Science Foundation, and the European Research Council (consolidator grant 616919-Ionology).

\bibliographystyle{tfp}

\end{document}